\documentclass[12pt]{article}

\usepackage[dvips]{graphicx}

\newcommand{\lsim}{\raisebox{-4pt}{$\,\stackrel{\textstyle
                                                         <}{\sim}\,$}}

\newcommand{\nn}{\nonumber}   
\newcommand{\be}{\begin{equation}}        
\newcommand{\ee}{\end{equation}}     
\newcommand{\ba}{\begin{eqnarray}}  
\newcommand{\ea}{\end{eqnarray}} 
\newcommand{\req}[1]{(\ref{#1})}
\def\={\,=\,}                      
\newcommand{\ci}[1]{\cite{#1}}
\def\gev{~{\rm GeV}}     
\def\mev{~{\rm MeV}}

\def\als{\alpha_{\rm s}} 
\def\LQCD{\Lambda_{\rm QCD}}  
\def\muF{\mu_F}             
\def\muR{\mu_R}       
\def\muO{\mu_{0}}             
\def\xb{\bar{x}}
\newcommand{\tw}{\textwidth}   
\def\vk{{\bf k}_{\perp}}     
\def\vbs{{\bf b}}               
\newcommand{\da}{{distribution amplitude}}  
\newcommand{\wf}{wave function} 

\begin{document}
\thispagestyle{empty}
\begin{flushright}
WU B 11-14 \\
September, 14 2011 \\[20mm]
\end{flushright}

\begin{center}
{\Large\bf The photon to pseudoscalar meson transition form factors} \\
\vskip 10mm

P.\ Kroll \footnote{Email:  kroll@physik.uni-wuppertal.de}
\\[1em]
{\small {\it Fachbereich Physik, Universit\"at Wuppertal, D-42097 Wuppertal,
Germany}}\\
and\\
{\small {\it Institut f\"ur Theoretische Physik, Universit\"at
    Regensburg, \\D-93040 Regensburg, Germany}}\\
\end{center}
\vskip 5mm

\begin{abstract}
In this talk it is reported on an analysis \ci{kroll10} of the form factors
for the transitions from a photon to one of the pseudoscalar mesons $\pi^0,
\eta, \eta^\prime, \eta_c$ within the modified perturbative approach in which
quark transverse degrees of freedom are retained. The report is focused on
the discussion of the surprising features the new BaBar data exhibit, namely
the sharp rise of the $\pi\gamma$ form factor with the photon virtuality and
the strong breaking of flavor symmetry in the sector of pseudoscalar mesons. 
\end{abstract}


\section{Introduction}
\label{sec:introduction}
The recent measurements of the photon-to-pseudoscalar-meson transition form
factors by the BaBar collaboration \ci{babar09,babar10,babar11} has renewed
the interest in these form factors. The BaBar data ruined the believe that these
form factors are well understood in collinear QCD. The $\pi\gamma$ form factor
in particular, measured up to about $35\,\gev^2$ now, reveals a sharp rise
with the photon virtuality, $Q^2$. In fact, the scaled form factor 
$Q^2 F_{\pi\gamma}$ approximately increases $\propto\sqrt{Q}$ which is in
dramatic conflict with dimensional scaling and turned previous calculations
based on collinear factorization obsolete - a substantial increase of the form
factor is difficult to accommodate in fixed order perturbative QCD. 
For large $Q^2$ the form factor reads
\ba
Q^2F_{\pi\gamma}(Q^2)&=&\frac{\sqrt{2}f_\pi}{3}\,\int_0^1\,dx\,\Phi_\pi(x,\muF)\nn\\
                      &\times&\frac1{x}\,\Big[1 + \frac{\als(\muR)}{2\pi}{\cal
                           K}(x,\ln(Q^2)\Big]\,.
\label{eq:pQCD}
\ea  
The LO order result has been derived in \ci{lep79}, and the NLO result, ${\cal K}$,
in \ci{NLO}. In \req{eq:pQCD} $f_\pi$ is the familiar decay constant of the
pion,  $\muF$ and $\muR$ are the factorization and renormalization scales,
respectively. The \da{}, $\Phi_\pi$, or more generally that of a pseudoscalar
meson, $P (=\pi^0, \eta, \eta^\prime)$, possesses a Gegenbauer expansion
\ba
\Phi_P(x,\muF) &=& \Phi_{\rm AS} \Big[1 + \sum_{n=2,4,\cdots} a^p_n(\muO) \nn\\ 
              &\times& \Big(\frac{\als(\muF)}{\als(\muO)}\Big)^{\gamma_n}
                C_n^{3/2}(2x-1)\Big]\,,
\label{eq:distribution-amplitude}
\ea
where the evolution of the expansion parameters, $a^p_n$, from an initial scale,
$\muO$, to the factorization scale, $\muF$, is controlled by the anomalous 
dimensions $\gamma_n$ \ci{lep79}. With $\muF\to\infty$ the \da{} evolves in
the asymptotic form $\Phi_{\rm AS}=6x(1-x)$. Obviously, \req{eq:pQCD} only provides 
$\log{Q^2}$ effects arising from the running of $\als$ and the evolution of
the \da. Surprisingly, and this is in addition to the sharp rise of
$F_{\pi\gamma}$ a second puzzle, the other transition from factors measured by
the BaBar collaboration, namely $\eta\gamma$, $\eta^\prime\gamma$ and
$\eta_c\gamma$, do not exhibit an anomalous $Q^2$ dependence, they behave as
expected, see e.g.\ \ci{feldmann97,feldmann97b}.

Many theoretical papers have already been devoted to the interpretation of the
BaBar data on the $\pi\gamma$ form factor. Limitation of space only allows to 
mention a few of them. Thus, for instance, the flat \da{}, $\Phi_\pi\equiv 1$
has been proposed \ci{rad09,pol09} which, as is evident from \req{eq:pQCD}, 
necessitates a regularization prescription for the infrared singular
integral. This changes the asymptotic behavior of $Q^2F_{\pi\gamma}$ from the 
constant $\sqrt{2}f_\pi$ to $\propto \ln{Q^2}$. Other approaches base on
light-cone sum rules \ci{agaev10,bakulev} or consider instanton effects
\ci{dorokhov10}. Another possibility is to use $\vk$-factorization instead of
the collinear approach. Here, in this talk, it will be reported on an analysis 
\ci{kroll10} of the transition from factors on the basis of that factorization.

\section{The transition form factors in $\vk$-factorization}
\label{sec:vk}
\begin{figure}[t]
\begin{center}
\includegraphics[width=0.65\tw,viewport=135 550 578 653,clip=true]{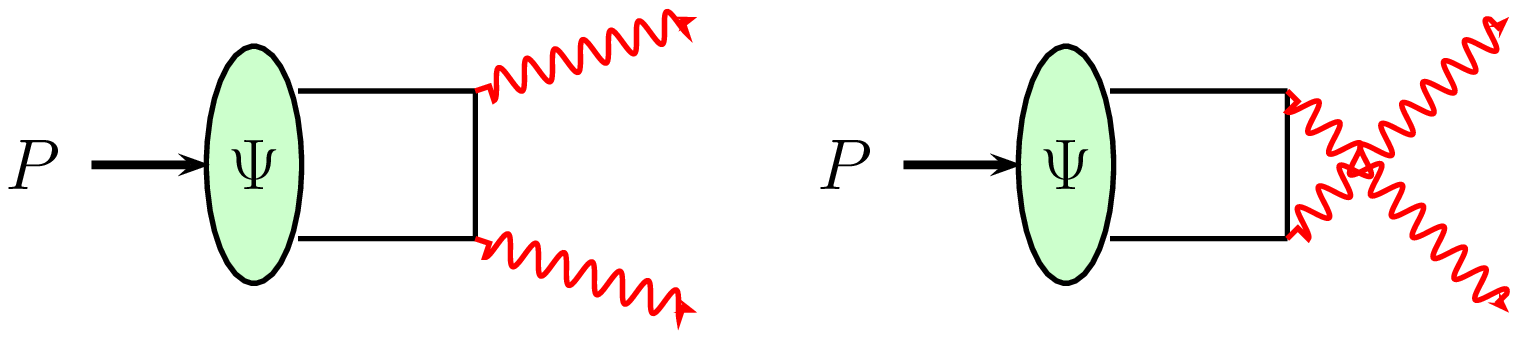}
\end{center}
\caption{Lowest order Feynman graphs for the $P\gamma$ transition form
  factor.}
\label{fig:LO-graphs}
\end{figure}

The basic idea of this approach is to retain the quark transverse 
degrees of freedom in the hard scattering. This however implies that quarks 
and antiquarks are pulled apart in the impact-parameter space, canonically 
conjugated to the transverse-momentum space. The separation of color sources 
is accompanied by the radiation of gluons. The corrections to the hard 
scattering process due to gluon radiation have been calculated in 
\ci{botts89} in axial gauge using resummation techniques and having recourse 
to the renormalization group. These radiative corrections comprising resummed 
leading and next-to-leading logarithms which are not taken into account by the 
usual QCD evolution, are presented in the form of a Sudakov factor, 
$e^{-S}$. This approach, termed the modified perturbative approach (MPA),
generates power corrections to the collinear result \req{eq:pQCD} which 
may explain the anomalous behavior of the BaBar data \ci{kroll10,li09}. 
According to \ci{raulfs95} the form factor for a transition from a photon to 
a pseudoscalar meson (P) reads
\ba
F_{P\gamma}(Q^2) &=& \int dx \frac{d^2\vbs}{4\pi}\,
\hat{\Psi}_P(x,-\vbs,\muF) \nn\\
&\times& \hat{T}_H^P(x,\vbs,Q,\muR) e^{-S(x, b, Q, \muR, \muF)}
\label{eq:FF-mpa}
\ea
within the MPA. Since the Sudakov exponent $S$ is given in the
impact-parameter space it is convenient to work in that space. In the
convolution formula \req{eq:FF-mpa} $\hat{T}_H$ is the Fourier transform of
the momentum-space hard scattering amplitude evaluated to lowest order 
perturbative QCD from the Feynman graphs displayed in Fig.\ \ref{fig:LO-graphs} 
\be
\hat{T}_H^P \= \frac{2\sqrt{6}\,C_P}{\pi}\,K_0(\sqrt{x}Qb)\,.
\label{eq:th}
\ee 
Here $C_P$ is a charge factor. For instance, for the pion it reads 
$C_\pi=(e_u^2-e_d^2)/\sqrt{2}$ where $e_a$ denotes the charge of a flavor-$a$
quark in units of the positron charge. The function $K_0$ denotes the modified
Bessel function of order zero.

The last item in \req{eq:FF-mpa} to be explained, is $\hat{\Psi}_P$, the
valence Fock state light-cone wave function of the meson P in the
impact-parameter space. In the original version of the MPA \ci{li92} this 
\wf{} is assumed to be just the meson \da{}. As argued in \ci{li92} 
the Sudakov factor, $e^{-S}$, can be viewed as the perturbatively generated 
transverse part of the \wf{}. For low and intermediate values of $Q^2$, 
however, the non-perturbative intrinsic $\vbs$- or $\vk$-dependence of the 
light-cone \wf{} cannot be ignored as has been pointed out in \ci{jakob93}. 
The inclusion of the transverse size of the meson extends considerably the 
region in which the perturbative contribution to the form factor can be 
calculated. As in \ci{raulfs95,jakob93} the \wf{} is
parameterized as
\be
\hat{\Psi}_P(x,\vbs,\mu_F) = 2\pi
\frac{f_P}{\sqrt{6}}\,\Phi_P(x,\muF)\,
                            \exp{\Big[-\frac{x\xb b^2}{4 \sigma_P^2}\Big]}\,.
\label{eq:Gaussian}
\ee

In the MPA the impact parameter, $\vbs$, which represents the transverse 
separation of quark and antiquark, acts as an infrared cut-off parameter. 
Thus, $1/b$ in the Sudakov exponent marks the interface between the 
non-perturbative soft momenta which are implicitly accounted for in the 
hadron \wf, and the contributions from soft gluons, incorporated in a 
perturbative way in the Sudakov factor. Obviously, the infra-red cut-off 
serves at the same time as the gliding factorization scale
\be
 \muF \= 1/b
\ee
to be used in the evolution of the \wf. In accord with this interpretation the
entire Sudakov factor is continued to zero whenever $b > 1\LQCD$. In this 
large-$b$ region the wavelength of the radiative gluon is larger than
$1/\LQCD$. Because of the color neutrality of the hadron such gluons cannot 
resolve the hadron's quark distribution; hence radiation is damped. Soft 
gluons with wavelength larger than $1/\LQCD$ are therefore to be excluded from 
perturbation theory; they have to be absorbed in the soft \wf. This implies
that the upper limit of the $b$-integral in \req{eq:FF-mpa} is $1/\LQCD$
instead of infinity. 

Radiative corrections with wavelengths between the infra-red cut-off and the
limit $\sqrt{2}/\xi Q$ yield to suppression through the Sudakov factor.
Gluons with even shorter wavelengths are regarded as hard ones 
which are considered as higher-order perturbative corrections of the hard 
scattering and, hence, are not part of the Sudakov factor but are included in
the hard scattering amplitude. The properties of the Sudakov function lead to 
an asymptotic damping of any contribution except those from configurations
with small quark-antiquark separations, i.e.\ for $\ln{Q^2}\to\infty$ the 
limiting behavior of the transition from factors in collinear factorization
emerges, for instance, $Q^2 F_{\pi\gamma}\to \sqrt{2}f_\pi$.  

In analogy to the case of the pion's electromagnetic form factor \ci{li92} 
the maximum of the longitudinal scale appearing in the scattering amplitude 
and the transverse scale
\be
\muR \= \max(\sqrt{x}Q,1/b)
\label{eq:renormalization}
\ee 
is chosen as the renormalization scale \ci{raulfs95}. Although to lowest
order there is no $\als$ in the hard scattering amplitude for the $P\gamma$ 
transition form factor, it nevertheless depends on $\mu_R$. Indeed, as 
discussed above, the Sudakov factor comprises the gluonic radiative corrections 
for scales between $1/b$ and $\xi Q/\sqrt{2}$. Hence, the latter scale
specifies the onset of the hard scattering regime. It is to be noted that
logarithmic singularities arising from the running of $\als$ and the evolution
of the \wf{} are canceled by the Sudakov factor.  

It can be shown \ci{BDK} that the Sudakov factor in combination with the hard
scattering amplitude provides a series of power suppressed terms which come
from the region of soft quark momenta ($x, 1-x \to 0$) and grow with the
Gegenbauer index $n$. This property leads to a strong suppression of the
higher-order Gegenbauer terms at low $Q^2$ implying that only the lowest few
Gegenbauer terms influence the transition form factor. With increasing $Q^2$
the higher Gegenbauer terms become gradually more important. At very large
$Q^2$ the evolution of the \da{} again suppresses all higher-order Gegenbauer
terms and the asymptotic limit of the transition form factor emerges. This is
to be contrasted with the collinear approximation \req{eq:pQCD} where, to
leading-order accuracy, the sum of all Gegenbauer terms (at a given
factorization scale) contributes to the form factor. The intrinsic transverse
momentum dependence embedded in the \wf{} also generates power suppressed
terms which are accumulated at all $x$ and do not depend on the Gegenbauer index. 
We note in passing that the suppression of higher-order Gegenbauer terms by soft
non-perturbative correction has also been observed within the framework of
light-cone sum rules \ci{agaev10}. 
  
\begin{figure}[t]
\begin{center}
\includegraphics[width=.65\tw, viewport=105 385 569 721,clip=true]
                {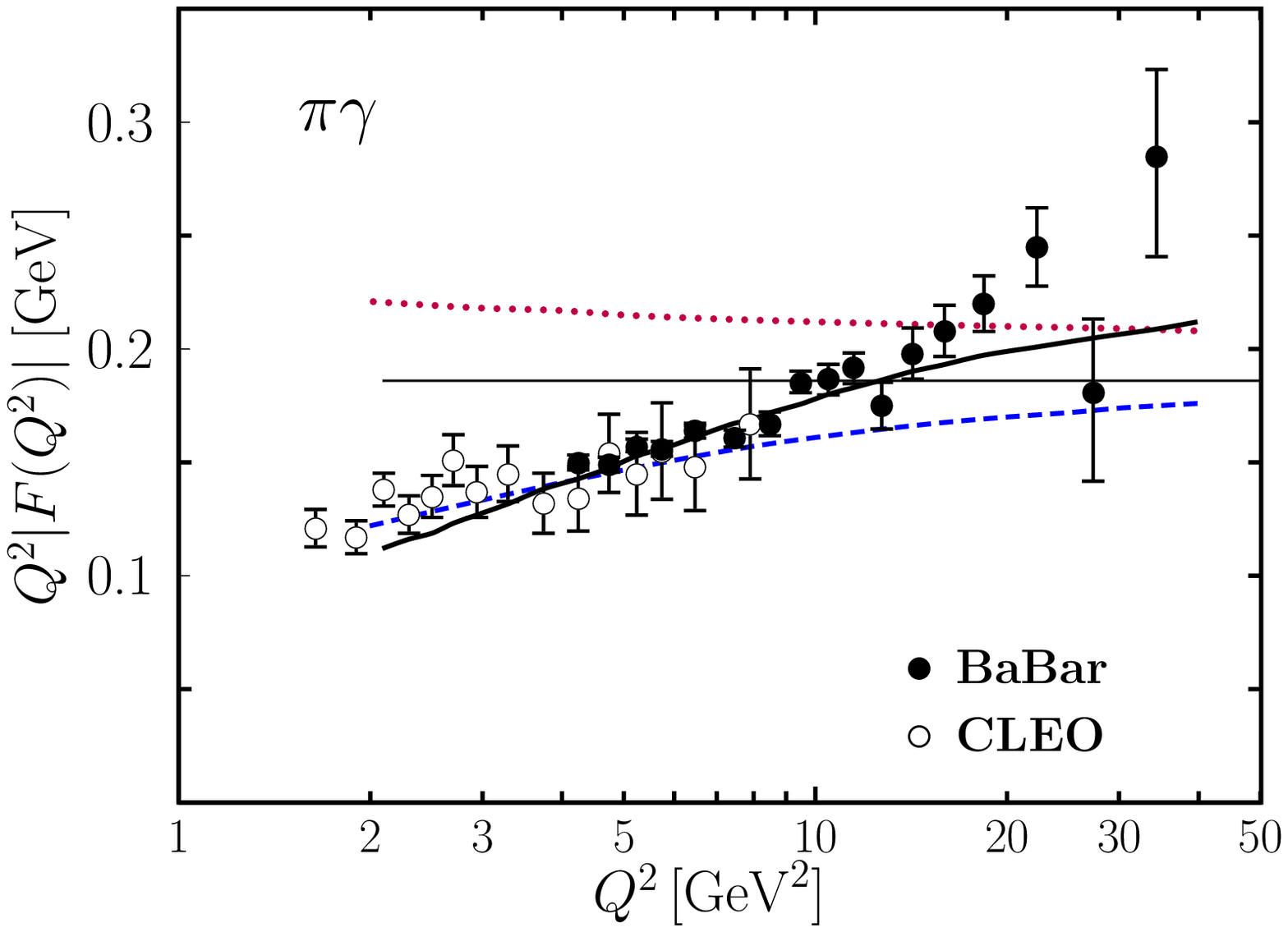} 
\end{center}
\caption{(Color online) The scaled $\pi\gamma$ transition form factor versus
  $Q^2$ evaluated from fit \req{eq:a4-fit} (solid line). The dashed line
  represents the result presented in \ci{raulfs95} which is obtained from the 
  asymptotic \da{} and $\sigma_\pi=0.861\,\gev^{-1}$. The dotted line is 
  obtained from collinear factorization to NLO accuracy. The thin solid line 
  indicates the asymptotic behavior. Data taken from \ci{babar09,CLEO}}
\label{fig:fit-a4}
\end{figure}

This feature of the MPA explains why the CLEO data on the $\pi\gamma$
transition form factor \ci{CLEO} are well described by the asymptotic 
\da{} as shown in \ci{raulfs95} (with $\sigma_\pi=0.861\,\gev^{-1}$,
$\LQCD=200\,\mev$), see Fig.\ \ref{fig:fit-a4}. With the BaBar 
data \ci{babar09} at disposal which extend to much larger values of 
$Q^2$ and do exceed the asymptotic limit $\sqrt{2} f_\pi$ , higher 
Gegenbauer terms can no more be ignored; they are required for a 
successful description of the transition form factor. What can be 
learned about the higher-order Gegenbauer terms from the BaBar data will 
be discussed in the next section.

\section{Confronting with experiment}
\label{sec:experiment}
Now, having specified the details of the MPA, one can analyze the transition
form factor by inserting \req{eq:Gaussian} and \req{eq:distribution-amplitude} 
into \req{eq:FF-mpa} and fitting the Gegenbauer coefficients to experiment.
From a detailed examination of the data it becomes apparent that besides the 
transverse size parameter only one Gegenbauer coefficient can safely be 
determined. If more coefficients are freed the fits become unstable. The 
coefficients acquire unphysically large absolute values between 1 and 10 
and with alternating signs leading to strong compensations among the various
terms.
 
Let us begin with the analysis of the CLEO \ci{CLEO} and BaBar \ci{babar09}
data on the $\pi\gamma$ form factor. A reasonable fit to these data is obtained
by taking for the Gegenbauer coefficient $a_2^\pi$ the face value of a lattice 
QCD result  \ci{lattice}:
 \be
a_2^\pi(\muO)\=0.20
\label{eq:a2}
\ee
at the scale $\muO=2\,\gev$ (with $\LQCD=0.181\,\gev$ and the 1-loop
expression for $\als$) and fitting $a_4^\pi$ and the transverse size
parameter, $\sigma_\pi$, to the data for $Q^2>2.3\,\gev^2$. The resulting
parameters are
\ba
\sigma_\pi&=&0.40\pm 0.06\,\gev^{-1}\,, \nn\\
 a_4(\muO)&=&0.01\pm 0.06\,, 
\label{eq:a4-fit}
\ea
and $\chi^2=34.2$ which appears reasonable given that 28 data points are
included in the fit. A fit with just $\sigma_\pi$ and $a_2^\pi$ leads to
parameters in agreement with the lattice result \req{eq:a2} and the parameters
quoted in \req{eq:a4-fit}; the results for the form factor are practically 
indistinguishable from the first fit. The results for the $\pi\gamma$ form
factor of the fit \req{eq:a4-fit} are shown in Fig.\ \ref{fig:fit-a4}. At the 
largest values of $Q^2$ the fit seems to be a bit small as compared to the
BaBar data. Partially responsible for this fact are the large fluctuations the 
BaBar data exhibit; the fit compromises between all the data. For comparison
the fit presented in \ci{raulfs95} which is evaluated from the asymptotic
\da{} and $\sigma_\pi=0.861\,\gev$, is also shown in Fig.\ \ref{fig:fit-a4}. 
This result is obviously too low at large $Q^2$, it does not exceed the
asymptotic limit of the scaled form factor, $\sqrt{2} f_\pi$, in contrast to the
BaBar data and the fit \req{eq:a4-fit}. Also shown in Fig.\ \ref{fig:fit-a4} 
is a typical result of the collinear factorization approach to NLO accuracy 
taking $\mu_F=\mu_R=Q$ and the Gegenbauer coefficients specified in \req{eq:a2} 
and \req{eq:a4-fit}. Apparently the shape of that result is in conflict with
experiment. The difference between this result and the MPA result for the
same \da{} reveals the strength of the power corrections taken into account by
the MPA. It is to be stressed that, in the collinear factorization approach, 
a change of the values of the Gegenbauer coefficients or the addition of
further coefficients with the proviso that large negative coefficients are 
excluded, does not alter the shape but only the absolute value of the form 
factor. Thus, the $\pi\gamma$ transition form factor sets an example of an 
exclusive observable for which collinear factorization is insufficient for 
$Q^2$ as large as $35\,\gev^2$.    

In Sec.\ \ref{sec:vk} it is mentioned that the Sudakov factor can be viewed as
the perturbatively generated transverse part of the \wf{}. In the the original
version of the MPA \ci{li92}, applied to the electromagnetic form factor of
the pion, only this part of the \wf{} has been taken into account and any
intrinsic transverse momentum neglected. With this approximation however,
i.e.\ if the Gaussian in \req{eq:Gaussian} is replaced by 1, a food fit to 
the form factor data cannot be achieved, the results are too flat as compared
to the data and the minimal $\chi^2$ is 155. Hence, the $\vbs$ dependence of
the \wf{} is an important ingredient of the MPA as has been suggested in 
\ci{jakob93}.

\begin{figure}[t]
\begin{center}
\includegraphics[width=.65\tw, viewport=110 380 585 730,clip=true]{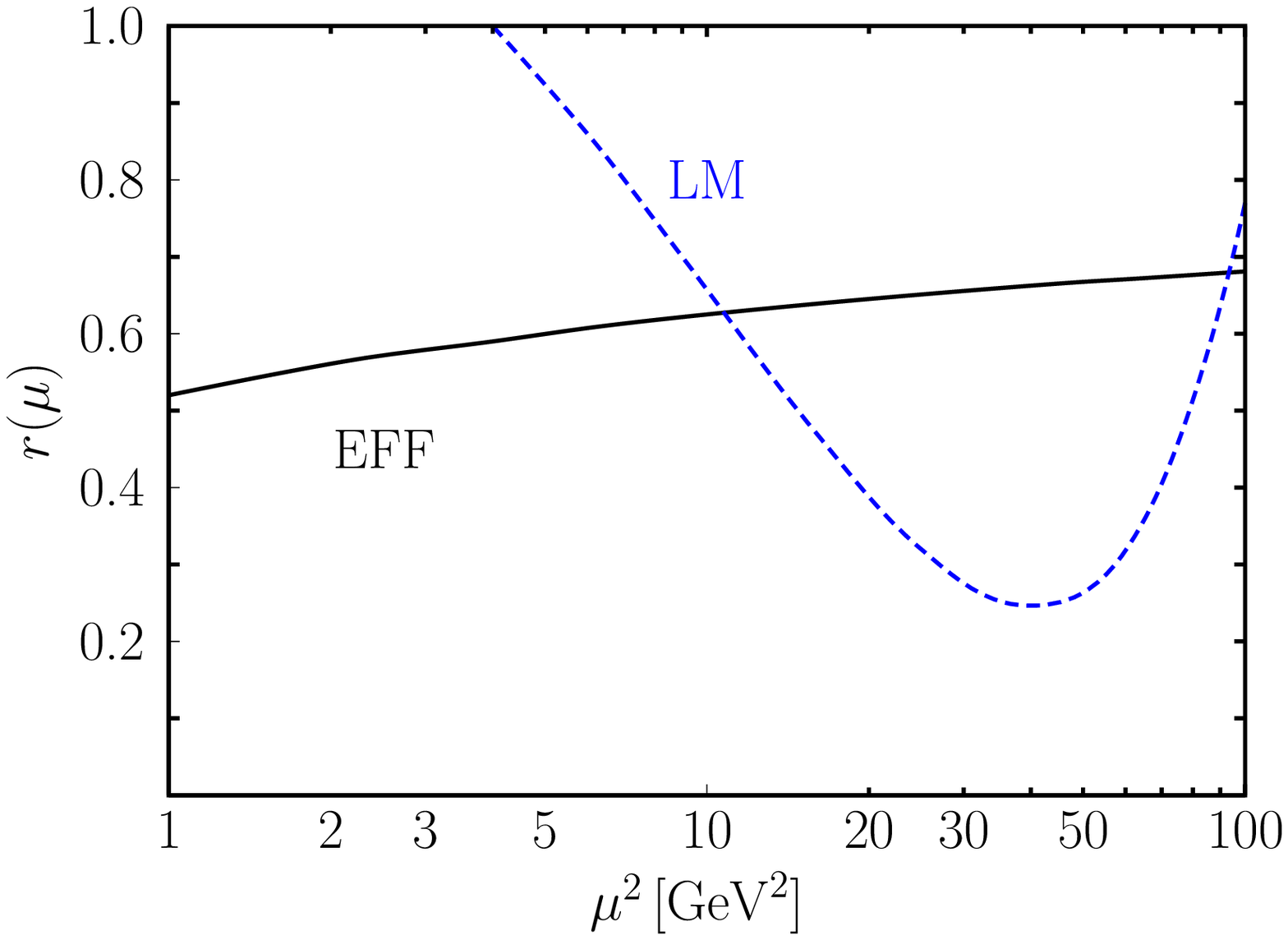} 
\end{center}
\caption{(Color online) The effective power (EFF) of the \da{} \req{eq:powerDA}
 compared to the power (LM) of the threshold factor in \ci{li09} at the scale $\mu$. }
\label{fig:power}
\end{figure}

Li and Mishima \ci{li09} also applied the MPA to the $\pi\gamma$ transition
form factor and achieved a reasonable fit to experiment. In contrast to
\ci{kroll10} the flat \da{}, $\Phi_\pi\equiv 1$, is used. It is combined with a 
Gaussian $b$-dependence as in \req{eq:Gaussian} in a kind of wave function. 
However, this product cannot be considered as a proper \wf{} in so far as it 
is not normalizable. It is furthermore argued in \ci{li09} that the flat \da{} 
is accompanied by a threshold factor that represents resummed double logs 
$\als \ln^2{x}$ and $\als \ln^2{(1-x)}$ arising from the end-point
singularities which occur for the flat \da{} in collinear factorization. 
The threshold factor combined with the flat \da{} can be viewed as an 
effective \da{} of the type
\be
 \Phi_\pi^r\= \frac{\Gamma(2+2r)}{\Gamma^2(1+r)}\,\big[x\xb\big]^{r}\,.
\label{eq:powerDA} 
\ee 
According to \ci{li09}, $r$ is about 1 for low $Q^2$ and small for   
$Q^2\simeq 35\,\gev^2$, see Fig.\ \ref{fig:power}. This particular
$Q^2$-dependence of the power $r$ generates the increase of the form factor 
required by the BaBar data \ci{babar09}: At low $Q^2$ the effective \da{} is
the asymptotic one implying small values of the transition form factor while, 
at $Q^2$ of about $35\,\gev^2$, the effective \da{} is close to the flat one 
and hence leads to much larger values of the form factor. 

Eq.\ \req{eq:powerDA} defines a family of power-like \da s. It includes the
limiting cases of the asymptotic \da{} for $r=1$ as well as the flat \da{}
for $r=0$. Also the square root \da{} proposed in \ci{bro07} belongs to this 
family. Expanding \req{eq:powerDA} upon the Gegenbauer polynomials and using 
its evolution one can show \ci{BDK} that, for $0\leq r(\muO)\leq 1$, the \da{} 
\req{eq:powerDA} approximately remains power-like under evolution, $r \to
r(\mu)$ over a large range of the scale. The power-like \da{} \req{eq:powerDA}
may be examined by fitting the transverse size parameter as well as the power 
$r(\muO)$ to the data on the $\pi\gamma$ form factor. One finds ($\chi^2\=34.4$)
\ba
\sigma_\pi&=& 0.40\pm 0.05\,\gev^{-1}\,,\nn\\ 
  r(\muO)&=&0.59\pm 0.06\,. 
\label{eq:powers}
\ea
The quality of this fit to the data on the $\pi\gamma$ form factor is 
similar to that presented in \ci{li09} and to \req{eq:a4-fit}. In Fig.\ 
\ref{fig:power} the power $r(\mu)$ for the fit \req{eq:powers} is 
compared to the scale dependence of the threshold factor used in \ci{li09} 
(in this work the power is set to unity for $\mu^2\lsim 4\,\gev^2$). As can be
seen from Fig.\ \ref{fig:power} the \da{} \req{eq:powerDA} exhibits the usual 
evolution behavior, it monotonically evolves into the asymptotic one, $\Phi_{\rm AS}$,
for $\mu\to\infty$. On the other hand, the scale dependence of the threshold
factor or the effective \da{} advocated for in \ci{li09}, is drastically
different. 

Let us now turn to the analysis of the $\eta\gamma$ and $\eta^\prime\gamma$
form factors. They can be expressed as a sum of the flavor-octet and
flavor-singlet contributions \ci{feldmann97} 
\be
F_{P\gamma} \= F^8_{P\gamma} +  F^1_{P\gamma}\,.
\ee
As is the case for the $\pi\gamma$ form factor the functions $F^i_{P\gamma}$
($i=1, 8$) are proportional to the constants $f_P^i$ assigned to the decays 
of meson $P$ through the SU(3)$_F$ octet or singlet axial-vector weak currents 
which are defined by the matrix elements
\be
\langle 0\mid J_{\mu 5}^i\mid P(p)\rangle \= i f_P^i\, p_\mu\,.  
\ee
Adopting the general parameterization \ci{kaiser}
\ba
f_\eta^8 &=& f_8 \cos{\theta_8}\,, \qquad f_\eta^1 \=- f_1 \sin{\theta_1}\,,\nn\\
f_{\eta^\prime}^8 &=& f_8 \sin{\theta_8}\,, \qquad 
f_{\eta^\prime}^1 \=\phantom{-} f_1 \cos{\theta_1}\,,
\ea
one can show \ci{FKS1} that on exploiting the divergences of the axial-vector
currents - which embody the axial-vector anomaly - the mixing angles,
$\theta_8$ and $\theta_1$, differ considerably from each other and from the
state mixing angle, $\theta$. In \ci{FKS1} the mixing parameters have
been determined:
\ba
f_8&=&1.26\, f_\pi\,, \qquad  f_1\=1.17\, f_\pi\,, \nn\\
\theta_8&=&-21.2^\circ\,, \qquad \theta_1\=-9.2^\circ\,.
\label{eq:mixing-par}
\ea 

Assuming particle-independent \ci{feldmann97,FKS1} \wf s, $\Psi^{8}$ and
$\Psi^{1}$, for the valence Fock states of the respective octet and singlet 
$\eta$ mesons and parameterizing them as in \req{eq:Gaussian} with decay
constants $f_8$ and $f_1$ instead of $f_\pi$,  one can cast the transition
form factors into the form
\ba
F_{\eta\gamma} &=& \cos{\theta_8}\,F^8 - \sin{\theta_1}\,F^1\nn\\
F_{\eta^\prime\gamma} &=& \sin{\theta_8}\,F^8 + \cos{\theta_1}\,F^1\,.
\label{FF-mixing}
\ea 
The charge factors in \req{eq:th} read (with $P=1,8$)
\ba
C_8&=&(e_u^2 + e_d^2 -2e_s^2)/\sqrt{6}\,, \nn\\
C_1&=&(e_u^2 + e_d^2 + e_s^2)/\sqrt{3}\,.
\ea 
The asymptotic behavior of the form factors is
\be
Q^2 F^8 \to \sqrt{\frac23} f_8\,, \qquad Q^2 F^1 \to \frac{4}{\sqrt{3}} f_1\,.
\label{octet-singlet-asymptotics}
\ee
The renormalization-scale dependence of the singlet-decay constant \ci{kaiser}
is omitted since the anomalous dimension controlling it is of order 
$\als^2$ and, hence, leads to tiny effects. 
 
For the singlet $\eta$ meson, $\eta_1$, there is also a glue-glue Fock component 
which, however, only contributes to NLO (or higher) of the hard scattering
\ci{shifman,grozin,passek}. In the MPA analysis the glue-glue Fock component 
does not contribute directly but only through the matrix of the anomalous
dimensions and the  mixing of the singlet quark-antiquark with the glue-glue 
\da{}. It is assumed in \ci{kroll10} that the Gegenbauer coefficients of the 
glue-glue distribution amplitude are zero at a low scale of order $1\,\gev$. 
Hence, the quark-antiquark singlet distribution amplitude practically evolves 
with the same anomalous dimensions as the octet distribution amplitude. 

The two form factors $F^8$ and $F^1$ can now be evaluated from \req{eq:FF-mpa} 
and \req{eq:th} in full analogy to the $\pi\gamma$ transition form factor. 
The data on $F^8$ and $F^1$ are extracted from the CLEO \ci{CLEO} and BaBar 
\ci{babar11} data using \req{FF-mixing}. As for the $\pi\gamma$ form 
factor only the transverse size parameter and one Gegenbauer coefficient for 
each \wf{} can be determined. The best fit is obtained with the parameters: 
\ba
\sigma_8&=& 0.84\pm 0.14\,\gev^{-1}\,, \nn\\
a_{2}^8(\muO)&=& -0.06\pm 0.06\,, \nn\\
\sigma_1&=& 0.74\pm 0.05 \,\gev^{-1}\,, \nn\\ 
a_{2}^1(\muO)&=& -0.07\pm 0.04\,.
\label{eta-fit}
\ea
\begin{figure}[t]
\begin{center}
\includegraphics[width=.65\tw, viewport=105 370 571 718,clip=true]
{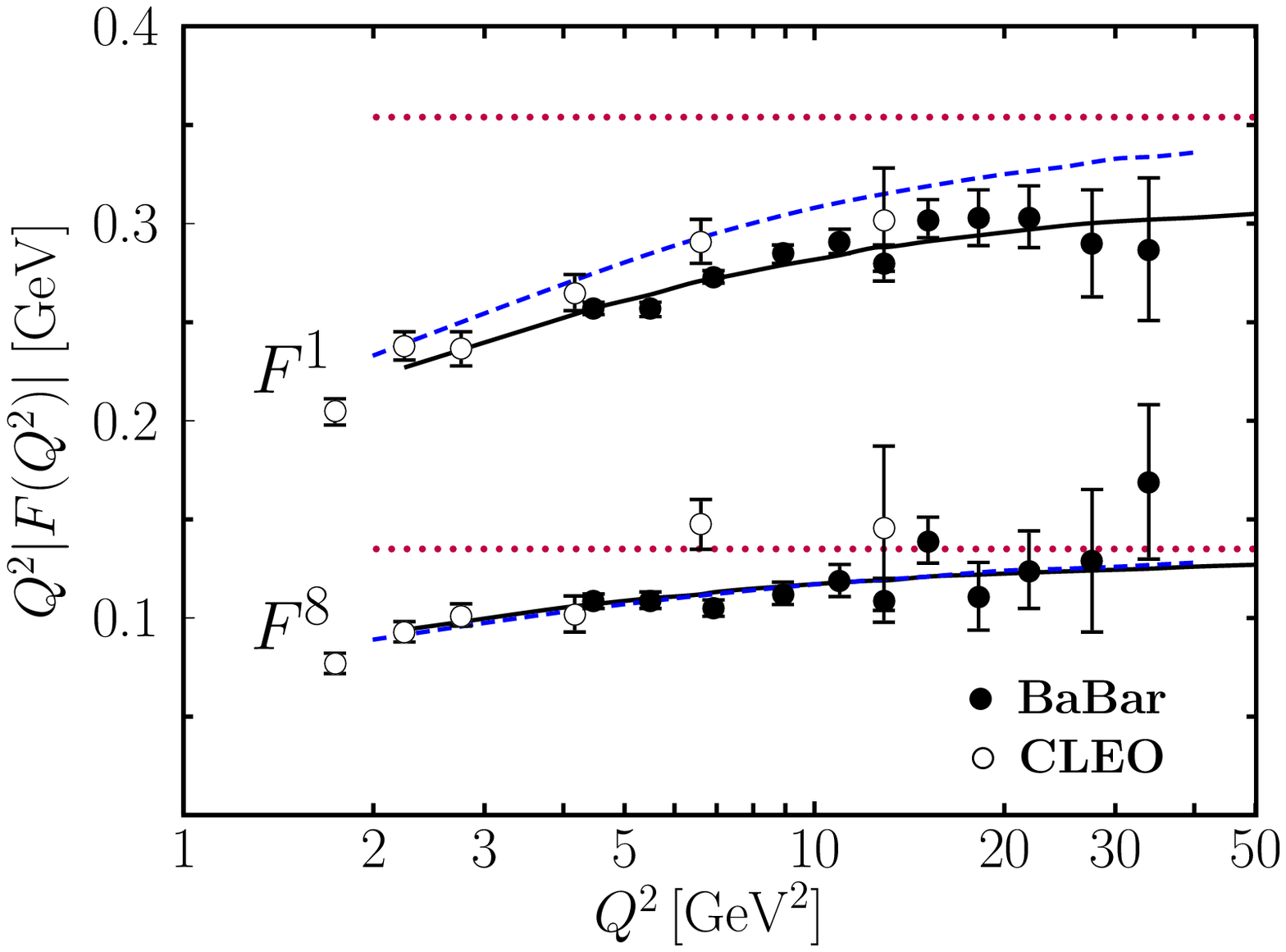}
  \end{center}
\caption{(Color online) The octet and singlet form factors. Dotted lines represent the
  asymptotic behavior \req{octet-singlet-asymptotics}, the dashed lines 
  the results obtained in \ci{feldmann97}. The solid lines represent the 
  new fit \req{eta-fit}. Data taken from \ci{CLEO,babar11}}
\label{fig:singlet-octet}
\end{figure}

\begin{figure}[t]
\begin{center}
\includegraphics[width=.45\tw, viewport=106 366 570 709,clip=true]
                                       {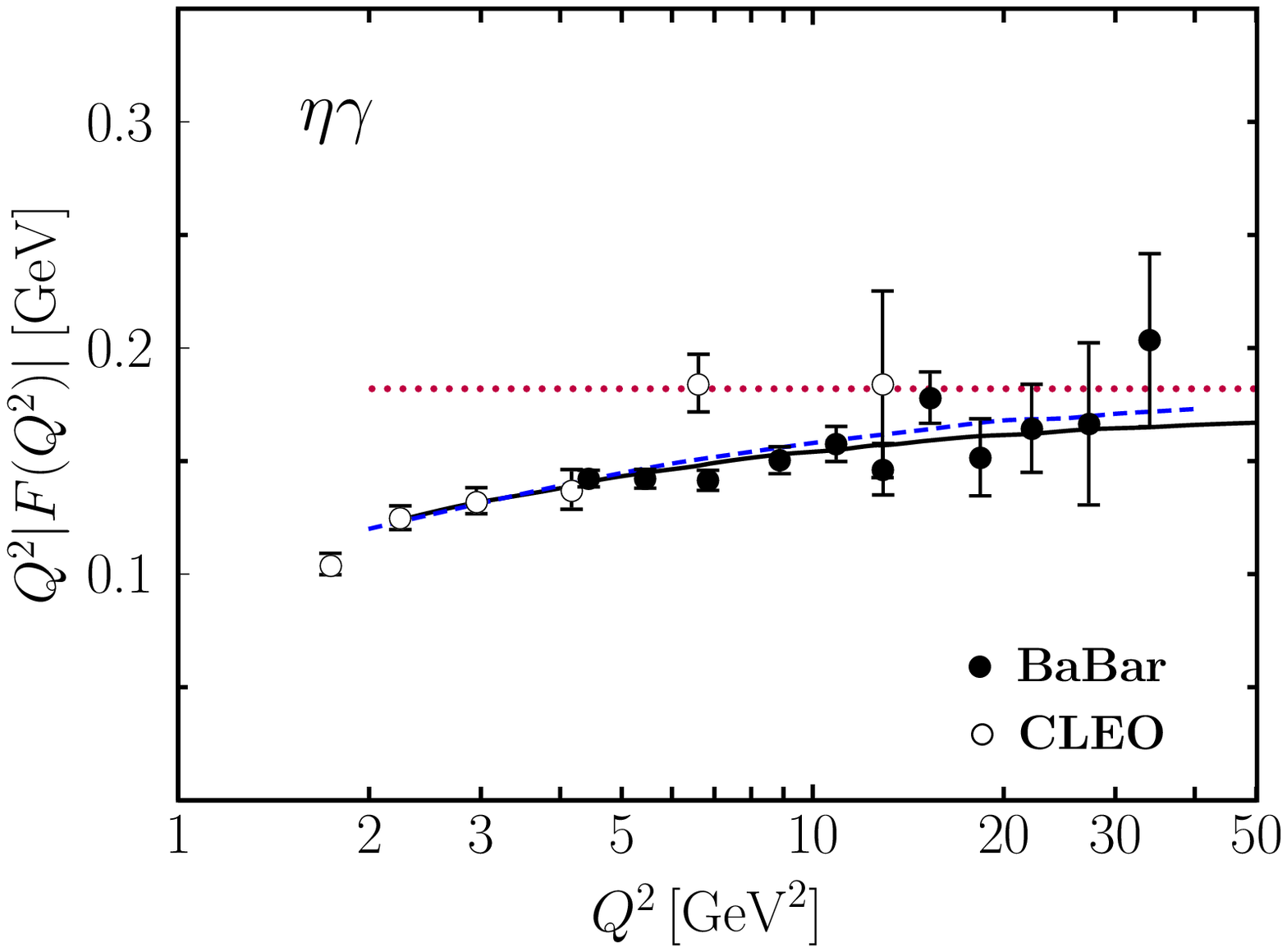}
\includegraphics[width=.45\tw, viewport=127 373 592 718,clip=true]
                                      {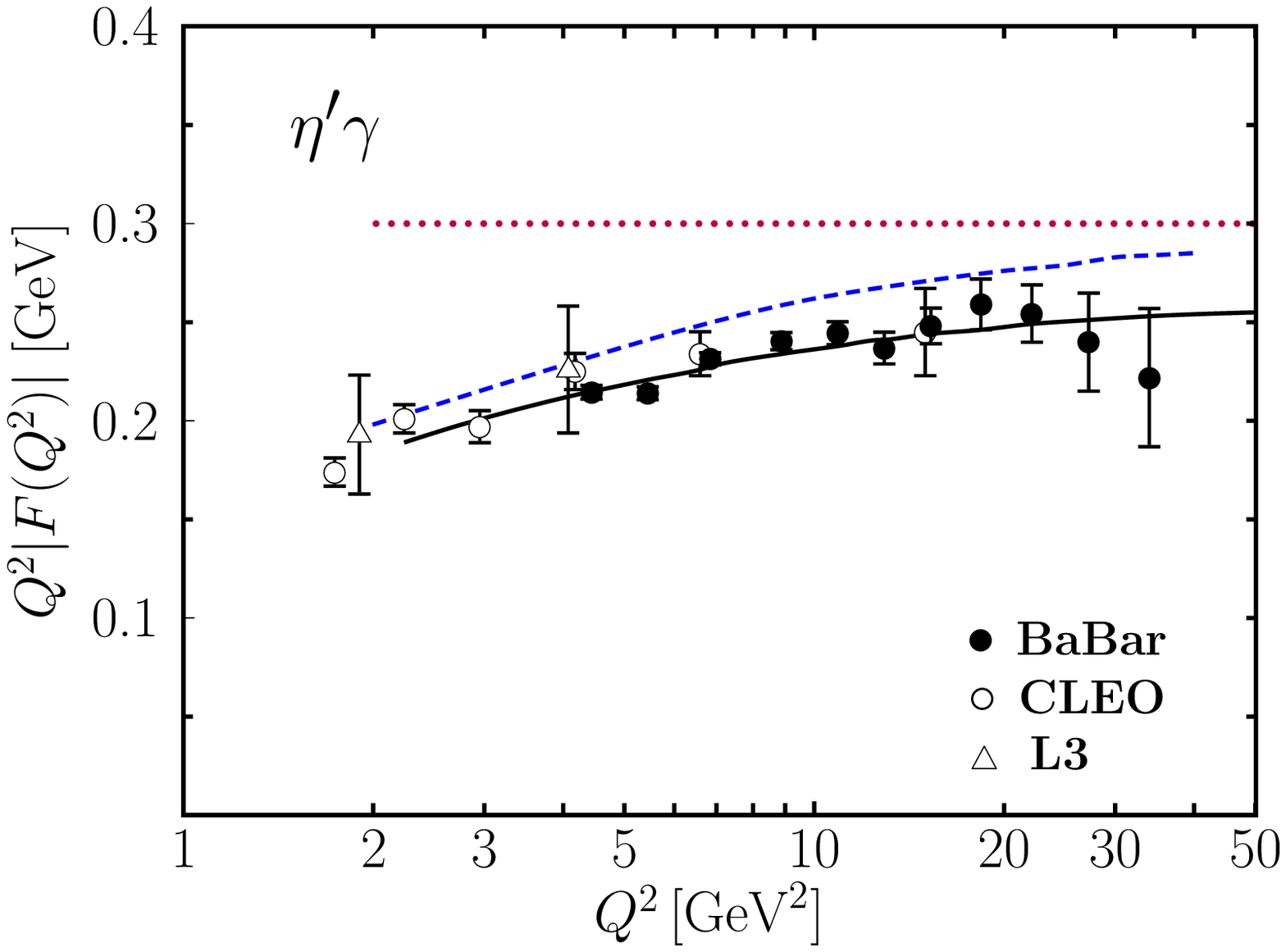}
\end{center}
\caption{(Color online) The scaled $\eta\gamma$ (left) and $\eta^\prime\gamma$ 
  (right) transition form factor versus $Q^2$. Data taken from \ci{CLEO,L3,babar11}.
  For notations refer to Fig.\ \ref{fig:singlet-octet}}
\label{fig:eta}
\end{figure}

The values of $\chi^2$ are 15.0 and 14.1 for the octet and singlet
cases, respectively (for 16 data points in each case). 
In Fig.\ \ref{fig:singlet-octet} the results of this fit are compared to the
data on $F^8$ and $F^1$. The quality of this fit is very good. In contrast to 
the $\pi\gamma$ case the data on both $F^8$ and $F^1$ lie below the asymptotic 
results \req{octet-singlet-asymptotics}. The combination of these two form 
factors into the physical ones according to \req{FF-mixing} leads to the
results shown in Fig.\ \ref{fig:eta}. Again very good agreement with the data 
is to be observed. Also shown in Figs.\ \ref{fig:singlet-octet} and 
\ref{fig:eta} are the results obtained in \ci{feldmann97} which have been 
evaluated from the asymptotic distribution amplitudes (with 
$\sigma_1=\sigma_{8}=0.861\,\gev^{-1}$). The octet as well as the $\eta\gamma$ 
form factors of \ci{feldmann97} are in very good agreement with experiment 
while the results for $F^1$ and the $\eta^\prime\gamma$ form factor are somewhat too large.  

Since the octet and singlet \wf s in \req{eta-fit} are very close to the
asymptotic one, the power corrections generated by the MPA are
small. Therefore, and in sharp contrast to the $\pi\gamma$ case, an analysis
of the $\eta\gamma$ and $\eta^\prime\gamma$ form factors within the collinear
factorization approach is also possible. In this case information on the
glue-glue Fock component of the $\eta_1$ may be extracted \ci{passek}. Even
with form factor data up to about $35\,\gev^2$ it is apparently not possible
to discriminate between logarithmic and power corrections.
 
The analysis of the $\eta\gamma$ and $\eta^\prime\gamma$ form factors
performed in \ci{kroll10} relies on the mixing parameters \req{eq:mixing-par}
determined in \ci{FKS1,FKS2}. Mixing of the $\eta$ and $\eta^\prime$ mesons has
been frequently investigated. Although in most cases not the same set of
processes as in \ci{FKS1,FKS2} has been analyzed the mixing parameters found
are in reasonable agreement with \req{eq:mixing-par}  within occasionally
large errors (see the discussion in \ci{kroll05}). An exception is
\ci{escribano} where the parameters markedly differ from \req{eq:mixing-par}. 
However, as pointed out in \ci{teryaev}, the parameters quoted in
\ci{escribano} are in conflict with the transition form factors.  

The treatment of the $\eta_c\gamma$, also measured by the BaBar collaboration
\ci{babar10}, differs from that of the three other cases. There is a second large
scale in addition to the virtuality of one of the photons, namely the mass of
the $\eta_c$ ($M_{\eta_c}$) or that of the charm quark ($m_c$). It cannot be
neglected in a perturbative calculation in contrast to the case of the light 
mesons where quark and hadron masses do not play a role. Despite this the 
$\eta_c\gamma$ form factor is to be calculated from \req{eq:FF-mpa} but the 
hard scattering amplitude to lowest order perturbative QCD reads \ci{feldmann97b}
\be
T_H \= \frac{4\sqrt{6}\, e_c^2}{xQ^2 + (1+4 x(1-x)) m_c^2+\vk^2}\,.
\label{eq:etac-hsa}
\ee
The symmetry of the problem under the replacement of $x$ by $1-x$ is
already taken into account in \req{eq:etac-hsa}. Due to the involved second 
large scale in the problem the $\eta_c\gamma$ form factor can be calculated 
even at $Q^2=0$.
 
The light-cone \wf{} of the $\eta_c$ is parameterized as in \req{eq:Gaussian}.
Following \ci{feldmann97b,bsw} the \da{} is chosen as  
\be
\Phi_{\eta_c} \= N(\sigma_{\eta_c}) x\xb 
   \exp{\Big[- \sigma^2_{\eta_c}M_{\eta_c}^2\frac{(x-1/2)^2}{x\xb}\Big]}
\ee
where $N(\sigma_{\eta_c})$ is determined from the usual requirement $\int_0^1
dx \Phi_{\eta_c}(x)=1$. The \da{} exhibits a pronounced maximum at $x=1/2$ and
is exponentially damped in the end-point regions. It describes an essentially 
non-relativistic $c\bar{c}$ bound state; quark and antiquark approximately 
share the meson's momentum equally. In the hard scattering
amplitude the charm quark mass occurs while in the \da{} the meson mass is
used. This property of the latter \da{} is a model assumption which
contributes to the theoretical uncertainty of the results. In the sense of 
the non-relativistic QCD \ci{HQET} $2m_c$ and $M_{\eta_c}$ are equivalent. 
In \req{eq:FF-mpa} the Sudakov factor can be set to 1 in the 
case at hand since it is mainly active in the end-point regions (see 
the discussion in Sect.\ \ref{sec:vk}) which are already strongly damped 
by the $\eta_c$ \wf. Even the $\vk$ dependence of the hard scattering
amplitude plays a minor role. The evolution behavior of the $\eta_c$ \da{} is
unknown in the range where $Q^2$ is of order of $M_{\eta_c}^2$ and is therefore 
ignored here. Consequently, also the running of the charm quark mass is 
omitted. 

The normalization of the $\eta_c\gamma$ transition form factor is fixed 
by its value at $Q^2=0$ which is related to two-photon decay width. 
However, this decay width is experimentally not well known \ci{PDG}. It is 
therefore advisable to normalize the form factor by its value at $Q^2=0$ all
the more so since the recent BaBar data \ci{babar10} are also presented this
way. Doing so the perturbative QCD corrections at $Q^2=0$ to the
$\eta_c\gamma$ transition form factor which are known to be large
\ci{barbieri}, are automatically included. Also the  $\als$ corrections for 
$Q^2\lsim M_{\eta_c}^2$ \ci{shifman} cancel to a high degree in the ratio 
$F_{\eta_c\gamma}(Q^2)/F_{\eta_c\gamma}(0)$. Even at $Q^2=10\,\gev^2$ their
effect is less than $5\%$, see the discussion in \ci{feldmann97b}. 
The uncertainties in the present knowledge of the $\eta_c$ decay constant 
do also not enter the predictions for this ratio.

\begin{figure}[t]
\begin{center}
\includegraphics[width=.65\tw, viewport=100 351 569 701,clip=true]{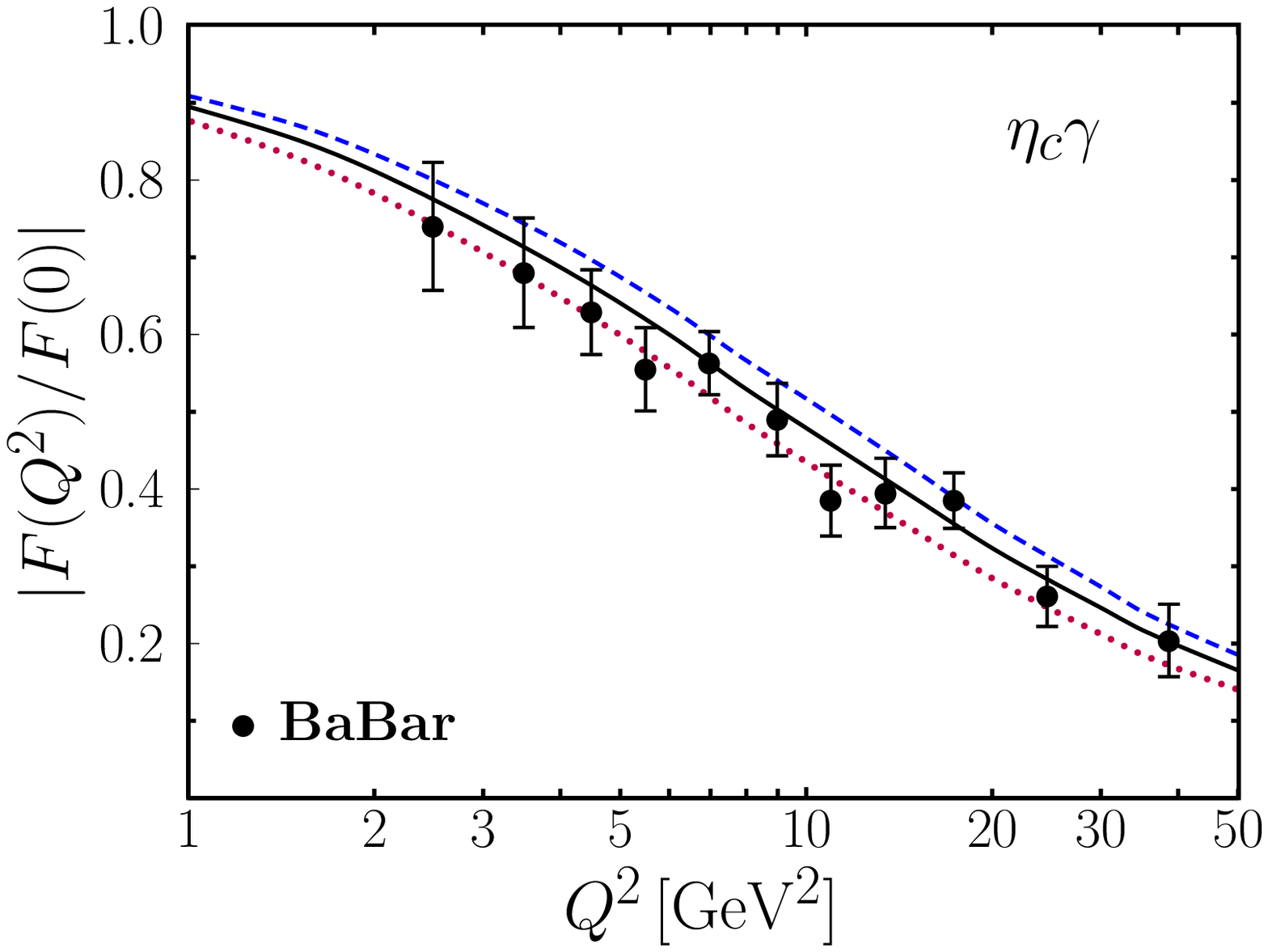}
 \end{center}
\caption{(Color online) The $\eta_c\gamma$ form factor scaled by its value at 
  $Q^2=0$. Data taken from \ci{babar10}. The solid (dotted) line represents
  the results of a calculation with the values of the parameters quoted in 
  \req{eq:etac-wf} (with $m_c=1.26\,\gev$). The dashed line is the prediction 
  given in \ci{feldmann97b}}
\label{fig:etac}
\end{figure}

The recent Babar data on the $\eta_c\gamma$ form factor \ci{babar10} are shown
in Fig.\ \ref{fig:etac}. The
predictions given in \ci{feldmann97b} which have been evaluated from 
$m_c=M_{\eta_c}/2$, are about one standard deviation too large 
but with regard to the uncertainties of the theoretical calculation, as for 
instance the exact value of the mass of the charm quark, one can claim 
agreement between theory and experiment. A little readjustment of the value of 
the charm quark mass improves the fit. Thus, with the parameters   
\be
m_c \=1.35\,\gev, \qquad \sigma_{\eta_c}\=0.44\,\gev^{-1}\,,
\label{eq:etac-wf}
\ee
a perfect agreement with experiment is achieved as is to be seen in Fig.\
\ref{fig:etac}. For comparison there are also shown results  evaluated from 
$m_c=1.21\,\gev$. 

\begin{figure}[t]
\begin{center}
\includegraphics[width=.65\tw, viewport=102 380 574 727,clip=true]
                                   {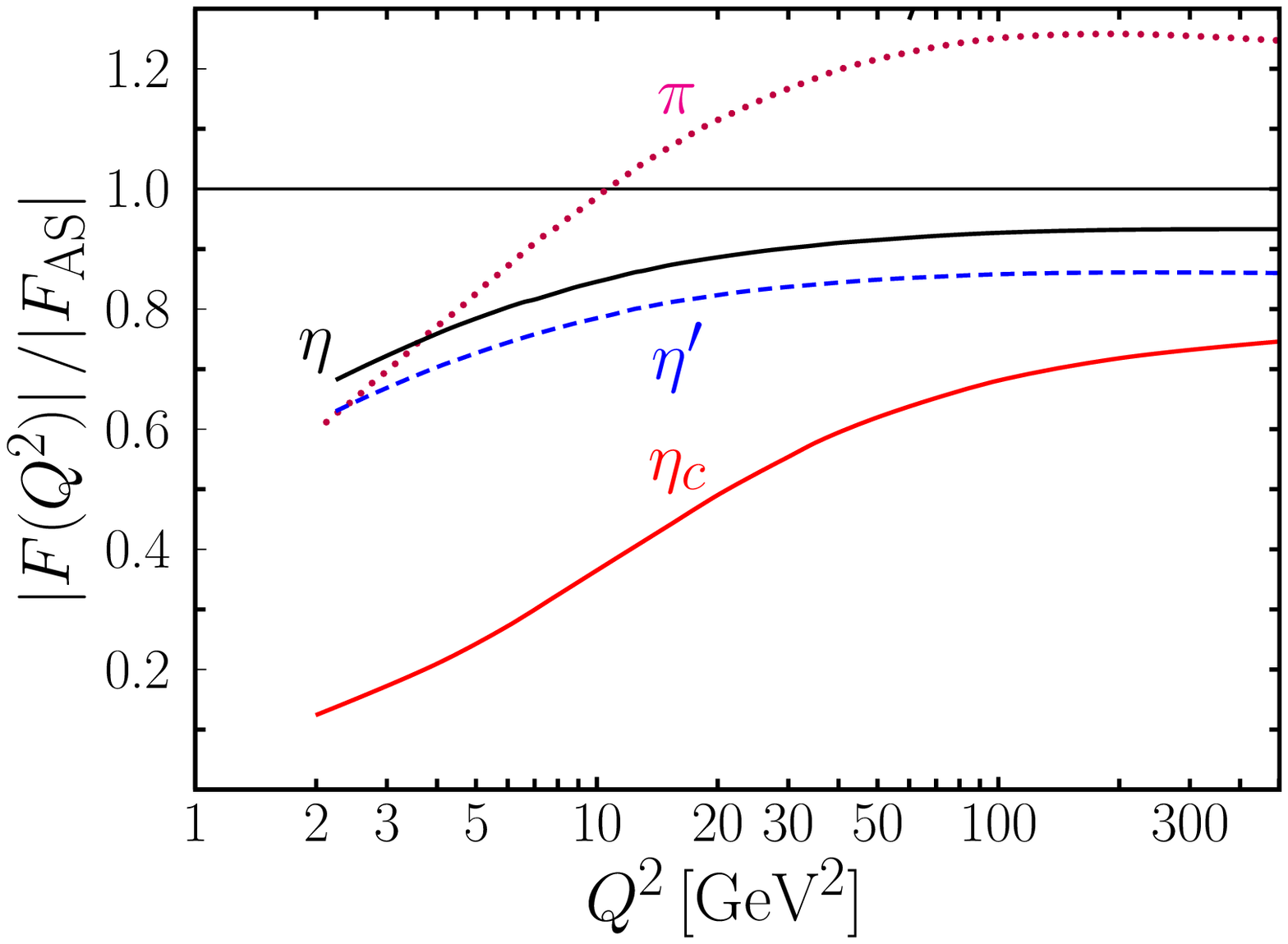}
\end{center}
\caption{(Color online) The $P-\gamma$ transition form factors scaled by the
  corresponding asymptotic behavior, versus $Q^2$. The thick solid (dashed,
  dotted, thin solid) line represents the case of the $\eta$ ($\eta^\prime$,
  $\pi$, $\eta_c$) meson. Parameters are taken from fit \req{eq:a4-fit}, 
  \req{eta-fit} and \req{eq:etac-wf}} 
\label{fig:comp}
\end{figure}

The form factors scaled by their respective asymptotic behaviors are displayed
in Fig.\ \ref{fig:comp} for a large range of $Q^2$. Asymptotically 
they all tend to 1. The $\pi\gamma$ form factor approaches 1 from above, the 
other ones from below. The approach to 1 is a very slow process; even at 
$500\,\gev^2$ the limiting behavior has not yet been reached. It is also
evident from Fig.\ \ref{fig:comp} that, forced by the BaBar data, there are 
strong violations of SU(3)$_F$ flavor symmetry in the ground state octet of the 
pseudoscalar mesons at large $Q^2$. In other processes
involving pseudoscalar mesons, e.g.\ two-photon annihilations into pairs 
of pseudoscalar mesons \ci{DK09,BELLE}, such large flavor symmetry violations 
have not been observed. Below $8\,\gev^2$, i.e.\ in the range of the CLEO
data, flavor symmetry breaking is much milder. The $\eta_c\gamma$ transition 
form factor which is also shown in Fig.\ \ref{fig:comp}, behaves  
different - the large charm-quark mass slows down the approach to the
asymptotic limit
\be
Q^2 F_{\eta_c\gamma} \to \frac{8f_{\eta_c}}{3}\,.
\ee  

\section{Summary}
In this talk it is reported on an analysis of the form factors for the
transitions from a photon to a pseudoscalar meson \ci{kroll10}. The analysis
is performed within the MPA which bases on $\vk$ factorization.  
In combination with the hard scattering kernel the Sudakov suppressions 
which are an important ingredient of the MPA and which represents radiative
corrections in next-to-leading-log approximation summed to all orders of 
perturbation theory, lead to a series of power suppressed terms which are
accumulated in the soft regions. Since these corrections grow with the
Gegenbauer index the transition form factors are only affected by the few 
lowest Gegenbauer terms of the \da{}, the higher ones do practically not 
contribute. How many Gegenbauer terms are relevant depends on the range of 
$Q^2$ considered: In the $Q^2$ range covered by the CLEO data \ci{CLEO} 
($<10\,\gev^2$) it suffices to use just the asymptotic \da{} in order to 
fit the CLEO data. With the BaBar data \ci{babar09,babar11} at disposal, 
covering the unprecedented large range $4\,\gev^2 < Q^2 < 35\,\gev^2$, the 
next or the next two Gegenbauer terms have to be taken into account or, 
turning the argument around, can be determined from an analysis of the data 
on the transition form factors. Indeed this is what has been done in
\ci{kroll10}. From this analysis it turned out that for the case of the pion 
a fairly strong contribution from $a_2$ is required by the data while for the 
$\eta$ and $\eta^\prime$ much smaller deviations from the asymptotic \da{} are 
needed. For these cases the results from a previous calculation within the MPA 
\ci{feldmann97} are already in fair agreement with the BaBar data, nearly 
perfect for the $\eta$, slightly worse for the $\eta^\prime$. Comparing the 
$\pi\gamma$ form factor with the $\eta\gamma$ or more precisely the 
$\eta_8\gamma$ one, one observes a strong breaking of flavor symmetry in the 
ground-state octet of the pseudoscalar mesons.  In other processes involving 
pseudoscalar mesons such large flavor symmetry violations have not been 
observed. With regard to the theoretical importance of the transition form 
factors, in particular the role of collinear factorization a remeasurement, 
e.g. by the BELLE collaboration, would be highly welcome. 

{\bf Acknowledgements}
It is a pleasure to thank Stanislav Dubnicka and Suzanna Dubnickova for the
kind invitation to the conference on Hadron Structure held in Tatranska Strba 
(Slovakia) and for the hospitality extended to the authors. 
This work is supported in part by the BMBF, contract number 06RY258.\\

\end{document}